\documentstyle[aps,preprint]{revtex}

\newcommand{\thH}{\mbox{$\theta_H$}}
\newcommand{\thc}{\mbox{$\theta_c$}}
\newcommand{\thz}{\mbox{$\theta_0$}}
\newcommand{\epsn}{\mbox{$\epsilon$}}

\newcommand{\ben}{\begin{equation}}
\newcommand{\een}{\end{equation}}

\newcommand{\taubar}{\mbox{$\bar{\tau}$}}
\newcommand{\delbar}{\mbox{$\bar{\delta}$}}

\begin{document}
\draft
\title{Quantum Nucleation in a Ferromagnetic Film
Placed in a Magnetic Field at an Arbitrary Angle}

\author{Gwang-Hee Kim$^*$}
\address{Department of Physics, Sejong University, Seoul 143-747
  , Korea}
\date{Received \hspace*{10pc}}
\maketitle

\begin{abstract}
We study the quantum nucleation in a thin ferromagnetic film
placed in a magnetic field at an arbitrary angle.
The dependence of the quantum nucleation and
the temperature of the crossover from thermal 
to quantum regime on the  direction and
the strength of the applied field are presented.
It is found that the maximal value of the rate and that of
the crossover temperature are obtained at a some angle
with the magnetic field, not in the direction of the
applied field opposite to the initial easy axis.\\
\end{abstract}             

\thispagestyle{empty}

\pacs{PACS numbers: 75.45.+j, 73.40.Gk, 75.30.Gw, 75.50.Gg\\
\\
\\
\\
\\
\\
\noindent
$*$ e-mail: gkim@phy.sejong.ac.kr}

Consider a ferromagnetic film of thickness $z_t$ and
its plane is perpendicular to the easy axis 
determined by
the magnetocrystalline anisotropy energy
depending on the crystal symmetry. Since the 
magnetic anisotropy energy has a time reversal symmetry
in the absence of an external magnetic field,
${\bf M}$ has at least two easy directions. By applying
a magnetic field ${\bf H}$ in a direction 
between perpendicular and opposite
to the initial easy axis along which the system is magnetized,
the symmetry of two equivalent orientations of ${\bf M}$ is
broken.  
According to the classical theory of nucleation,\cite{fre}
the switching of the magnetization occurs for a magnetic
field greater than the critical field $H_c$ at which
the spins of a saturated ferromagnet cease to be aligned. 
Besides the magnetic anisotropy energy $E_a$, the exchange
energy $E_{ex}$ and the magnetic dipole-dipole 
interation $E_{dd}$ determine the switching process of the
system. In this paper we simplify the system with small
$E_{dd}$ compared with $E_a$ and $E_{ex}$.\cite{chu,fer} 
At finite temperature, even for the field less than
$H_c$, there is a nucleation due to thermal fluctuation
whose rate is proportional to $\exp(-U/k_B T)$ 
where $U$ is an energy barrier related to the magnetic
anisotropy energy, the external magnetic field
and the exchange energy. At a 
temperature low enough to neglect the thermal activation
process, the magnetic nucleation which is independent of
temperature, may occur due to quantum tunneling. For the
dynamical situation, it is important to include the effect 
of the environment on the quantum tunneling rate caused
by phonons,\cite{gar}
nuclear spins,\cite{gar1} and Stoner
excitations and eddy currents in metallic magnets.\cite{tat}
However, many studies have shown that while these
coupling might be crucial in macroscopic quantum coherence,
they are not strong enough to make the
quantum tunneling unobservable.
 
The question of the quantum nucleation were studied by 
Privorotskii\cite{pri} who suggested the WKB exponent
based on the dimensional analysis. Later, Chudnovsky and
Gunther\cite{chu} proposed the quantum nucleation in 
a ferromagnetic
thin film at zero temperature by using the instanton
technique.
In their work it was considered the system with the external
magnetic field applied in the opposite direction to the
initial easy axis,
and presented the rate of quantum nucleation which allows
experimental observation with a sufficient probability.
In this paper we shall extend their considerations to the 
system with the magneic field applied 
in a range of angles $\pi/2 < \thH \leq \pi$ 
where $\thH$ is the
angle between the initial easy axis ($z$-axis) and the
external magnetic field, which, in its method and spirit,
is the generalization of Ref. \cite{mig} to problems with
a non-uniform rotation of $\bf{M}$.

Using the parameter defined to be $\epsn=1-H/H_c$,
at zero temperature the nucleation 
exponent $B_Q \propto \epsn^{3/4}$ in a wide range of
angles, not $\epsn^{1/2}$ which was 
obtained by Chudnovsky and Gunther
for $\thH=\pi$.\cite{chu} The crossover
temperature $T_c$ for quantum nucleation
to thermal activation has a weak dependence
on $\epsn$, $T_c \propto \epsn^{1/4}$
in a wide range of angle\cite{mig}, compared with
$\epsn^{1/2}$ at $\thH=\pi$.\cite{chu}
Also, it is found that
the direction of $\bf{H}$ which favors the quantum
nucleation lies in the range of angle
$\pi/2 < \thH < \pi$,
not $\pi$. In this paper we shall calculate
instantons in the range of angles
$\pi/2 + O(\sqrt{\epsn}) < \thH < \pi - O(\sqrt{\epsn})$
at $T=0$ based on the imaginary time path integral method
and their contribution to the quantum nucleation
exponent $B_Q$, and discuss the $\thH$ dependence of
$B_Q$ and the crossover temperature $T_c$ from
quantum nucleation to thermally assisted process.\\

Our considerations begin by writing the transition
amplitude in the imaginary time $\tau=it$ from 
$-\tau_{I}/2$ to $\tau_{I}/2$ based on the spin
coherent state path integral formalism
\ben
 \langle \hat{z}|\exp(-\tau_{I}H/\hbar)|\hat{z}
 \rangle =\int D [{\bf M}({\bf r},\tau)] 
 \exp(-S_E/\hbar),
\label{amp}
\een
where $\hat{z}$ is taken to be
an initial easy direction of the magnetization,
and
$S_E$ is the Euclidean action which includes
the Euclidean Lagrangian density $L_E$ as 
\ben
 S_E=\int d\tau \int d^3 {\bf r} L_E.
\een
Since
the magnitude of the total magnetization
$\bf{M}$ is frozen up at sufficiently
low temperature, the direction of the
total magnetization becomes
the only dynamical
variable which we are interested in. 
In this situation 
the measure of the path 
integral $D [{\bf M}({\bf r},\tau)]$ in 
Eq. (\ref{amp}) is 
replaced by $D \Omega$ as
\begin{equation}
D \Omega =\lim_{\epsilon\rightarrow 0}
\prod_{k=1}^N\Bigl( {2J+1\over 4\pi}\Bigr)\sin{\theta_k}\,
d\theta_k\, d\phi_k\, ,
\label{mea}
\end{equation}
where $\epsn=max(\tau_{k+1}-\tau_k)$ and $J=M_0/\hbar \gamma$.
Here $\gamma=g \mu_B /\hbar$, with $g$ being the g-factor,
$\mu_B$ the Bohr magneton, $M_0$ the magnitude of the
magnetization. 
Using the standard instanton method, 
the rate of quantum nucleation has 
the relation
\begin{equation}
 \Gamma_Q 
 \ \ \ \propto
 \ \ \ \exp(-S_E/\hbar),
\label{rate}
\end{equation}
and its 
Euclidean action for this problem is given by
\begin{equation}
  S_E=\int\{i \frac{M_0}{\gamma}
  [1-\cos \theta({\bf r},\tau)] \frac{d \phi({\bf r},\tau)}{d \tau}
  + E[{\bf M}({\bf r},\tau)]\} d \tau d^3 {\bf r} . \label{se}
\end{equation}
The first term in Eq. (\ref{se})
is the topological
Wess-Zumino term\cite{los},
and the second term is the energy density which is composed of the
magnetic
anisotropy energy, the exchange energy and the energy
given by an external magnetic field ${\bf H}$.
So, the energy taken  into account
can be written as
\ben
 E[{\bf M}({\bf r},\tau)]=E_a+E_{ex}-{\bf M} \cdot {\bf H}.
 \label{tote}
\een
We notice that from $\delta S_E=0$ the action (\ref{se}) produces
the classical
equation of motion\cite{gil}
whose solution is called instanton,\cite{lan} and
describes the 
$1 \oplus 1$ dimensional dynamics
in the Hamiltonian formulation,
which consists of the canonical coordinates
$\phi$ and $p_{\phi}=1- \cos \theta$.\cite{joh,kim}

In this paper we study the biaxial symmetry whose magnetic anisotropy
energy is given by
\ben
 E_a=K_1 (\alpha_{1}^{2}+\alpha_{2}^{2})+K_2 \alpha_{2}^{2},
\een
where $K_1$ and $K_2$ are the parallel and
transverse anisotropy constants, and
$\alpha_1, \alpha_2$, and $\alpha_3$
the directional cosines. The
exchange interaction is expressed as\cite{fre}
\ben
 E_{ex}={1 \over 2} C [(\nabla \alpha_1)^2+
        (\nabla \alpha_2)^2+(\nabla \alpha_3)^2]
\een
where $C$ is an exchange constant. Applying
the magnetic field in the $xz$-plane\cite{mig}
and expressing 
$\alpha_1$, $\alpha_2$ and $\alpha_3$ in terms of 
the fields $\theta({\bf r},\tau)$ 
and $\phi({\bf r},\tau)$
which are spherical coordinates of
${\bf M}$, the total energy
(\ref{tote}) is given by
\begin{eqnarray}
 E[\theta({\bf r},\tau),\phi({\bf r},\tau)]
  &=&K_1 \sin^2 \theta + K_2 \sin^2 \phi
  \sin^2 \theta 
  +{1\over 2}C[(\nabla \theta)^2+(\nabla \phi)^2 
  \sin^2\theta]
  \nonumber \\
  & &-H_x M_0 \sin \theta \cos \phi  -
  H_z M_0 \cos \theta + E_0 ,
\label{eq:toten}
\end{eqnarray}                                      
where $E_0$ is introduced to 
make $E(\theta, \phi)$ zero at the initial
orientation. As will be seen later, 
the effective
mass of the magnetic particle is inversely 
proportional to a linear combination of $K_2$
and $H_x$ while there is no exact analog
of the kinetic energy in the action (\ref{se}).
For this reason we need
either the transverse anisotropy
constant $K_2$ or the magnetic field $H_x$
transverse to the initial easy axis
for quantum nucleation.
Introducing the dimensionless
constants,
\begin{eqnarray}
 \bar{K}_2 = K_2/2K_1, \ \ \ \bar{C}
 =C/4K_1, \ \ \ \bar{H}_x
 = H_x/H_0, \ \ \
 \bar{H}_z = H_z/H_0,
 \label{eq:dimpar}
\end{eqnarray}   
where $H_0 = 2K_1/M_0$, we obtain 
the total energy (\ref{eq:toten}) 
written as
\begin{eqnarray}
 \bar{E}[\theta({\bf r},\tau),\phi({\bf r},\tau)]
 &=&\frac{1}{2} \sin^2 \theta
 +\bar{K}_2 \sin^2 \phi \sin^2 \theta 
 +\bar{C}[(\nabla \theta)^2+(\nabla \phi)^2 
 \sin^2\theta]
 \nonumber \\
 & & - \bar{H}_x
 \sin \theta \cos \phi  
 - \bar{H}_z \cos \theta + \bar{E}_0,
\label{eq:reden}
\end{eqnarray}
where $\bar{E}(\theta, \phi)=E(\theta, \phi)/2K_1$.
The plane given by $\phi =0$ is the easy plane, on which
$\bar{E}(\theta, \phi)$ is given by        
\begin{equation}
 \bar{E}(\theta,0)=\frac{1}{2} \sin^2 \theta
 -\bar{H} \cos (\theta- \theta_H ) + \bar{E}_0,
 \label{en0}
\end{equation}
where we used $\bar{H}_x=\bar{H} \sin \theta_H$ and
$\bar{H}_z=\bar{H} \cos \theta_H$ since ${\bf H}$ lies
in $xz$-plane.
Let us define $\theta_0$ to be the angle of metastable
state generated by the anisotropy energy and 
the external magnetic field, and
$\theta_c$ the angle at which the barrier vanishes by
the applied critical magnetic field $H_c$.
Then, 
$\theta_0$ is determined by 
$[d\bar{E}(\theta,0)/d\theta]_{\theta=\theta_0}=0$, 
and $\theta_c$ and $\bar{H}_c$ by 
$[d\bar{E}(\theta,0)/d\theta]_{\theta=\theta_c,\bar{H}=\bar{H}_c}=0$
and 
$[d^2\bar{E}(\theta,0)/d\theta^2]_{\theta=\theta_c,
\bar{H}=\bar{H}_c}=0$.
After a little calculations,  
the dimensionless critical field 
$\bar{H}_c$ and the critical angle $\theta_c$
are expressed as\cite{kim,mig}
\begin{eqnarray}
 \bar{H}_c &=&(\sin ^{2/3} \theta_H +
 |\cos \theta_H|^{2/3})^{-3/2}, \label{eq:hcb}\\
 \thc&=&{1\over2}\arcsin[{2|\cot\thH|^{1/3}\over
 1+|\cot\thH|^{2/3}}]. \label{eq:sthcb}
\end{eqnarray}
For, e.g.,
$\theta_H= 3\pi/4$ and $\pi$, we have
$\theta_c = \pi/4$ and 0,
respectively.
Simple analysis for Eq. (\ref{en0}) shows that,
since the height and the width of the barrier 
are proportional to the power of the parameter $\epsn$,
the value of $\epsn$ should be small in order to have
a large tunneling rate. 
The small value of $\epsn$ can be achieved by
tuning $H$ to $H_c$.
From now on we will consider the value of $\epsn$ to be
small, that is to say, $\epsn \ll 1$.
From the equations which
$\theta_0$ and $\theta_c$ satisfy,
we obtain the equation for $\eta(=\thc-\thz)$ as\cite{kim}
\begin{equation}
\sin(2 \theta_c) (\epsilon - \frac{3}{2} \eta^2)
- \eta \cos(2 \theta_c)(2 \epsilon - \eta^2)=0.
\label{eq:etaeq}
\end{equation}
Simple calculations
show that $\eta$ is of the order of $\sqrt{\epsn}$.
Thus
the order of magnitude of the second term in
Eq. (\ref{eq:etaeq}) is smaller than that
of the first term by $\sqrt{\epsn}$ and the 
value of
$\eta$ is determined by the first term
which leads to 
$\eta \simeq \sqrt{2 \epsilon /3}$.
However, when $\thH$ is very close to
$\pi/2$ or $\pi$, $\sin (2 \theta_c)$
becomes almost zero, as shown in Fig. \ref{figsth},
and the first term is much smaller than the
second term in Eq. (\ref{eq:etaeq}).
Therefore
the value of $\eta$ is obtained from
the second term
when $\thH \simeq \pi/2$ or $\pi$, 
which leads to
$\eta \simeq \sqrt{2 \epsilon}$
for $\theta_H \simeq \pi/2$ and $\eta \simeq 0$
for $\theta_H \simeq \pi$
from the detailed analysis of
the equations which 
$\theta_0$ and $\theta_c$ satisfy.
Since the first term in (\ref{eq:etaeq})
is dominant in the range of value $\thc$ which 
satisfies $\tan(2\thc) > O(\sqrt{\epsn})$,
$\eta \simeq \sqrt{2 \epsn /3}$
is valid for 
$\pi/2 +O(\sqrt{\epsn}) < \thH < \pi -O(\sqrt{\epsn})$
by usnig
Eq. (\ref{eq:sthcb}).
As is shown in Fig. \ref{etaval},
this is checked by performing the
numerical calculation for the equations which
$\thz$ and $\thc$ satisfy.
From Eq. (\ref{eq:reden})
we obtain 
an approximate form of
$\bar{E} (\theta, \phi)$
as
\begin{eqnarray}
 \bar{E}(\delta, \phi)&=&\bar{K}_2 \sin^2 \phi \sin^2
 (\theta_0 + \delta) + \bar{H}_x \sin (\theta_0+\delta)
 (1-\cos \phi) \nonumber \\ 
 & & +\bar{C}[(\nabla \delta)^2+(\nabla \phi)^2 
 \sin^2 (\thz+\delta)]
 +\bar{E}_1(\delta),
\label{eq:totdel}
\end{eqnarray}
where $\bar{E} (\theta, \phi)$ is written as
$\bar{E}(\delta, \phi)$ by introducing
a small variable $\delta =
\theta - \theta_0$, and  $\bar{E}_1(\delta)$ is a function of
only $\delta$ given by
\begin{equation}
 \bar{E}_1 (\delta)=\frac{1}{4} \sin (2 \theta_c)(3 \delta^2
                 \eta - \delta^3)+\frac{1}{2} \cos
                 (2 \theta_c)[\delta^2 (\epsilon -
                 \frac{3}{2} \eta^2)+\delta^3 \eta -
                 \frac{\delta^4}{4}].
\label{eq:edelta}
\end{equation}
As previously considered for $\eta$
in Eq. (\ref{eq:etaeq}), even though the
$\cos(2\thc)$-term
in Eq. (\ref{eq:edelta}) looks smaller by a factor of
$\eta$ which is
of the order of $\sqrt{\epsilon}$, the second term can not be neglected
near $\theta_H = \pi/2$ and $\pi$ because $\sin ( 2 \theta_c)$
is almost zero for these regions of $\thH$. 
In order to evaluate 
the order of magnitude of each term in the action
(\ref{se}) and simplify the calculations for
$\epsn \ll 1$, it is convenient to use
new scaled variables 
\begin{eqnarray}
 \tilde{\tau} = \epsn^{\alpha \over 2} \  \omega_0 \tau, \ \ \
 \tilde{{\bf r}}= \epsn^{\beta \over 2} \  {\bf r}/r_0, \ \ \ 
 \bar{\delta}=\delta/\sqrt{\epsn}, \ \ \ r_0=\sqrt{2\bar{C}}, \ \ \
 \omega_0 = 2 \gamma K_1 /M_0,
 \label{eq:dimpar1}
\end{eqnarray}
where it is noted that
$r_0$ is the thickness of the domain wall in
a bulk ferromagnet.
Then, the Euclidean
action (\ref{se}) becomes
\begin{eqnarray}
  S_E[\delta(\tilde{{\bf r}},\tilde{\tau}),
  \phi(\tilde{{\bf r}},\tilde{\tau})]&=& 
  \hbar J \epsn^{-{\alpha+3\beta \over 2}} r^{3}_{0} 
  \int d \tilde{\tau} d^3 \tilde{{\bf r}}
   \{ i \epsn^{\alpha \over 2}
   [1-\cos (\theta_0 + \sqrt{\epsn}\bar{\delta})]
  \frac{\partial \phi}{\partial \tilde{\tau}} \nonumber \\
  & & +\bar{K_2} \sin^2 \phi \sin^2 (\thz+\sqrt{\epsn}\bar{\delta})
  +\bar{H}_x \sin(\thz+\sqrt{\epsn}\bar{\delta})(1-\cos\phi)
  \nonumber \\
  & & +{1\over 2} \epsn^{1+\beta}(\tilde{\nabla} \bar{\delta})^2
  +{1\over 2} \epsn^{\beta}\sin^2 (\thz+\sqrt{\epsn}\bar{\delta})
  (\tilde{\nabla} \phi)^2 
  \nonumber \\
  & & + \frac{1}{4} \sin (2 \theta_c)\epsn^{3\over 2}(3 \delbar^2
  {\eta \over \sqrt{\epsn}} - \delbar^3)
  +\frac{1}{2} \cos (2 \theta_c) \epsn^2
  [\delbar^2 (1 - {3\eta^2 \over 2\epsn})
  +\delbar^3 {\eta \over \sqrt{\epsn}}
  -\frac{\delbar^4}{4}] \}.
\label{eq:newactn}
\end{eqnarray}
where the parameters
$\alpha$ and $\beta$ are fixed by the analysis of the
order of magnitude of each term in
the action (\ref{eq:newactn}) subject to the situation considered.\\

For $\pi/2 + O(\sqrt{\epsn}) < \thH < \pi - O(\sqrt{\epsn})$, 
the critical angle
$\thc$ and the angle of metastable state $\thz$
are in the range of angle from 0 to $\pi/2$, and 
$\eta \simeq \sqrt{2\epsn/3}$. 
In this situation, the action (\ref{eq:newactn})
becomes 
\begin{eqnarray}
  S_E[\delta(\tilde{{\bf r}},\tilde{\tau}),
  \phi(\tilde{{\bf r}},\tilde{\tau})]&=&
  \hbar J \epsn^{-{\alpha+3\beta \over 2}} r^{3}_{0}
  \int d \tilde{\tau} d^3 \tilde{{\bf r}}
   \{ -i \epsn^{\alpha+1 \over 2}
    \sin (\theta_0 + \sqrt{\epsn}\bar{\delta}) \ \phi \ 
  (\frac{\partial \delbar}{\partial \tilde{\tau}}) \nonumber \\
  & & +\bar{K_2} \sin^2 \phi \sin^2 (\thz+\sqrt{\epsn}\bar{\delta})
  +2\bar{H}_x \sin(\thz+\sqrt{\epsn}\bar{\delta})
  \sin^2({\phi \over 2})
  \nonumber \\
  & & +{1\over 2} \epsn^{1+\beta}(\tilde{\nabla} \bar{\delta})^2
  +{1\over 2} \epsn^{\beta}\sin^2 (\thz+\sqrt{\epsn}\bar{\delta})
  (\tilde{\nabla} \phi)^2
  \nonumber \\
  & & + \frac{1}{4} \sin (2 \theta_c)\epsn^{3\over 2}(3 \delbar^2
  {\eta \over \sqrt{\epsn}} - \delbar^3) \},
\label{eq:actn135}
\end{eqnarray}
where we performed the integration by part for the 
first term and neglected the total time derivative.
Since the last term in the action (\ref{eq:actn135})
makes the potential barrier in 
our problem, the order of magnitude of each term 
should be the same order as that of the last term.
Since the order of magnitude of each term is
$O(\epsn^{(\alpha+1)/2} \phi)$, $O(\sin^2\phi)$, 
$O(\sin^2(\phi/2))$, $\epsn^{1+\beta}$,
$O(\epsn^{\beta}\phi^2)$ and $\epsn^{3/2}$, respectively,
the small values of $\phi$ contribute to 
the path integral for $\epsn \ll 1$,
and its order of magnitude is expected to be
$\epsn^{3/4}$. Choosing $\alpha=\beta=1/2$, 
the order of magnitude of the term which includes
$(\tilde{\nabla} \phi)^2$ is $\epsn^2$ while 
those of the other terms are $\epsn^{3/2}$.
Thus, the fifth term in
Eq. (\ref{eq:actn135}) is higher and neglected.

Performing the Gaussian integration over $\phi$ in the
transition amplitude (\ref{amp})
and the measure (\ref{mea}), the
remaining integral is of the form
\ben
 \int D[\delta(\tilde{{\bf r}},\tilde{\tau})]
  \exp(-S^{eff}_{E}/\hbar),
\label{path180}
\een
where from Eq. (\ref{eq:actn135})
the effective action is given by
\ben
  S^{eff}_{E}[\delta(\tilde{{\bf r}},\tilde{\tau})]=
   \hbar J  r^{3}_{0} \epsn^{1 \over 2}
   \int d \tilde{\tau} d^3 \tilde{{\bf r}}
   [  {1\over 2} M
   ({\partial \delbar \over \partial \tilde{\tau}})^2
   +{1\over 2} (\tilde{\nabla} \delbar)^2
   + \frac{1}{4}\sin(2\thc) (\sqrt{6}\delbar^2-\delbar^3)].
\label{actn135red}
\een
Here the effective mass in the reduced dimension is given by
\ben
 M={\sin \thc \over \bar{H}_x+2\bar{K}_2 \sin \thc}.
\een
Introducing the variables 
$\bar{\tau}=\tilde{\tau}\sqrt{\sin (2\thc)/M}$ and 
$\bar{\bf{r}}=\tilde{\bf{r}}\sqrt{\sin (2\thc)}$,
we simplify 
the effective action (\ref{actn135red}) to be
\ben
  S^{eff}_{E}[\delta(\bar{\bf r},\bar{\tau})]=
   \hbar J  r^{3}_{0} \epsn^{1 \over 2} 
   {\sqrt{M} \over \sin(2 \thc)}
   \int d \bar{\tau} d^3 \bar{\bf r}
   [  {1\over 2} 
   ({\partial \delbar \over \partial \bar{\tau}})^2
   +{1\over 2} (\bar{\nabla} \delbar)^2
   + \frac{1}{4} (\sqrt{6}\delbar^2-\delbar^3)],
\label{actn135red1}
\een

In case of a small magnetic particle of volume $V \ll r^{3}_{0}$,
$\delbar$ does not depend on the space $\bar{\bf r}$, 
which leads to
the action 
\ben
  S^{eff}_{E}[\delta(\bar{\tau})]=
   \hbar J   \epsn^{5 \over 4} V \sqrt{M \sin(2\thc)}
   \int d \bar{\tau}
   [  {1\over 2}
   ({d \delbar \over d \bar{\tau}})^2
   + \frac{1}{4} (\sqrt{6}\delbar^2-\delbar^3)].
\label{actn135redu}
\een
The corresponding
classical trajectory is given by
\ben
 \bar{\delta}_{cl}(\bar{\tau}) = \frac{\sqrt{6}}
 {\cosh^2( ({3 \over 32})^{1/4} \bar{\tau})}, 
 \label{delcl135}
\een
and its classical action is found to be\cite{kim}
\begin{equation}
 S_{cl}= (\hbar J) \frac{16 \times 6^{1/4}}{5}
 \epsilon^{5/ 4} \frac{|\cot \theta_H|^{1/6}}
 {\sqrt{1+\frac{K_2}{K_1}(1+|\cot \theta_H|^{2/3})}}.
 \label{eq:scl135}
\end{equation}

The geometry of the problem in this paper is the
thin film of thickness $z_t$ less than the size
$r_0/\epsn^{1/4}$ of the critical nucleus and its
plane is perpendicular to
the initial easy axis.
After performing the integration over $\bar{z}$-variable,
the action (\ref{actn135red1}) becomes
\ben
  S^{eff}_{E}[\delta(\bar{\bf r},\taubar)]=
  \hbar J  r^{2}_{0} \  z_t \ \epsn^{3 \over 4}
  \sqrt{M \over \sin(2\thc)} 
  \int d \taubar d^2 \bar{\bf r}
   [  {1\over 2} (\partial_{\taubar} \delbar)^2
     +{1\over 2} (\partial_{\bar{x}} \delbar)^2
     +{1\over 2} (\partial_{\bar{y}} \delbar)^2
  + \frac{1}{4}  (\sqrt{6}\delbar^2-\delbar^3)],
\label{actn135red3}
\een
At zero temperature
the classical solution of the action (\ref{actn135red3})
has $O(3)$ symmetry in $\taubar$ and $\bar{\bf r}$.
Therefore,
$\delbar$ is a function of $q$
where $q=(\bar{x}^2+\bar{y}^2+\taubar^2)^{1/2}$,
in which
the effective action (\ref{actn135red3}) becomes
\ben
  S^{eff}_{E}[\delta(\bar{\bf r},\taubar)] =
  4 \pi \hbar J  r^{2}_{0}  z_t \epsn^{3 \over 4}
  f(\thH)
  \int^{\infty}_{0} dq q^2
  [ {1\over 2} ({d \delbar \over dq})^2
  + \frac{1}{4}  (\sqrt{6}\delbar^2-\delbar^3)],
\label{actn135red4}
\een
with the angular dependence
\ben
  f(\thH)={ 1+|\cot\thH|^{2/3} \over
  \sqrt{2}|\cot\thH|^{1/6}
  \sqrt{1-\epsn+{K_2 \over K_1}(1+|\cot\thH|^{2/3})} }.
\label{fth}
\een
The corresponding  classical equation of motion
satisfies
\ben
 {d^2 \delbar \over dq^2}
 + {2 \over q} {d \delbar \over dq}
 = {\sqrt{6} \over 2}\delbar - {3 \over 4} \delbar^2.
\label{ins135}
\een
This equation can be 
solved by the numerical approach,
whose results are
similar to the ones in Ref. \cite{chu}.
The maximal value of $\delbar$ is 6.85
at $\taubar=0$ and $\bar{r}=0$. Using the results, we 
obtain the rate of quantum nucleation given by
\begin{equation}
 \Gamma_Q
 \ \ \ \propto
 \ \ \ \exp(-S_E/\hbar)
 = \exp[-105.2 J  r^{2}_{0} f(\thH)
   z_t \epsn^{3/4}].
\label{rate1}
\end{equation}

Since the period $\tau_I$ of the instanton becomes
zero for 
the temperature much larger than $T_c$, the instanton
solution becomes independent of $\taubar$. In this case
we have
the rate of thermal nucleation 
\ben
 \Gamma_T
 \ \ \ \propto
 \ \ \ \exp(-{U^{eff} \over k_B T})
\label{rateth135}
\een
where the effective height of the barrier
\ben
U^{eff}= 4 \pi \  K_1 r^{2}_{0} \  z_t \epsn 
 \int^{\infty}_{0} dq q
 [ {1\over 2} ({d \delbar \over dq})^2
   +{\sqrt{6}\over 4} \delbar^2 - {1\over 4} \delbar^3].
\een
From the saddle point of the functional
the shape of thermal nucleus is determined by
\ben
 {d^2 \delbar \over d q^2}
 + {1 \over q}{d \delbar \over d q}
 -{\sqrt{6}\over 2} \delbar+{3 \over 4}\delbar^2=0.
\label{insth135}
\een
The solution is found by the numerical method whose
shape is similar to the one in Ref. \cite{chu}.
The maximal size of the thermal
nucleus is 3.9 at $\bar{\bf{r}}=0$. Using this result,
the effective height of the barrier becomes
\ben
 U^{eff}= 41.3 K_1 r^{2}_{0} z_t \epsn.
\een
Comparing this with Eq. (\ref{rate1}), we
obtain the approximate form of 
the temperature of the crossover from thermal
to quantum regime given by
\ben
 T_c \approx 0.22 {K_1  \epsn^{1/4} \over k_B J f(\thH)}.
\een
Here we notice that the angular and the field dependence of
$T_c$ coincides with the one of Ref. \cite{kim} and
with that of Ref. \cite{mig} in the limit of $\bar{K}_2=0$.
It is understood from the fact that 
$T_c \sim \hbar \omega_b/k_B$, where
the frequency $\omega_b$
of small oscillations around the minimum of the
inverted potential is proportional to
$\omega_0 \epsn^{1/4} / f(\thH)$ 
from Eq. (\ref{eq:dimpar1}) and the scaling 
relation between $\taubar$ and $\tilde{\tau}$.

Let us now consider the situation with the magentic field
opposite to the initial easy axis.
For $\thH=\pi$, we obtain $\thc = 0$ and $\eta = 0$ 
by using Eqs. (\ref{eq:sthcb}) and (\ref{eq:etaeq}).
In this case, noting that $\thz=0$, $\bar{H}_x=0$ and
$\epsn \ll 1$, the Euclidean
action (\ref{eq:newactn}) can be simplified to be
\begin{eqnarray}
  S_E[\delta(\tilde{{\bf r}},\tilde{\tau}),
  \phi(\tilde{{\bf r}},\tilde{\tau})]&=&
  \hbar J \epsn^{-{\alpha+3\beta \over 2}} r^{3}_{0}
  \int d \tilde{\tau} d^3 \tilde{{\bf r}}
   [  {i \over 2} \epsn^{1+{\alpha \over 2}} \
   \delbar^2 \
  \frac{\partial \phi}{\partial \tilde{\tau}}
  +\bar{K_2} \epsn \  \delbar^2 \sin^2 \phi
  \nonumber \\
  & & +{1\over 2} \epsn^{1+\beta}(\tilde{\nabla} \bar{\delta})^2
  +{1\over 2} \epsn^{1+\beta}\delbar^2
  (\tilde{\nabla} \phi)^2
  + \frac{1}{2}  \epsn^2 (\delbar^2-\frac{\delbar^4}{4})].
\label{actn180}
\end{eqnarray}

In order to obtain the
quantum tunneling under the barrier,
we need the second term which
includes the transverse component of 
the anisotropy energy $E_a$.
Without $K_2$ in $E_a$, $M_z$, as a quantum operator, commute
with $E_a$ and conserves. Thus, $K_2$ is responsible for
quantum transition at $\thH=\pi$.
Comparing the second term with the last term which
makes the potential barrier, only small values of $\phi$
contribute to the path integral because $\epsn \ll 1$.
Thus we expect the
order of magnitude of $\phi$ to be $\epsn^{1/2}$.
Choosing $\alpha=\beta=1$,
the order of magnitude of
each term except the fourth term 
in Eq. (\ref{actn180}) is $\epsn^2$ while
the fourth one is 
$\epsn^3$ which is higher and neglected.
Following the same procedure as we have done previously,
we obtained the WKB exponent $B_Q (T=0,\thH=\pi)$, 
the effective height of the barrier $U^{eff}(\thH=\pi)$ for
$T > T_c$ and the approximate form of the crossover
temperature $T_c$, which were presented by 
Chudnovsky and Gunther\cite{chu}.
Recently, Ferrera and Chudnovsky discussed the thermally
assisted quantum nucleation at $\thH=\pi$ and obtained
the temperature dependence of the WKB exponent
$B(T,\thH=\pi)$ in the range of temperature
$0 < T < 2 T_c$ by numerically solving the 
partial differential equation with period
$\hbar \omega \epsn^{1/2}/T$ where 
$\omega=(2\gamma/M_0)\sqrt{K_1 K_2}$.

As summarized in Table \ref{tab}, $B_Q$ is proportional to 
$\epsn^{3/4}$ for a range of angles
$\pi/2 + O(\sqrt{\epsn}) < \thH < \pi - O(\sqrt{\epsn})$,
while $B_Q \propto \epsn^{1/2}$ at $\thH=\pi$. By using
the interpolation method, we obtain $B_Q$ in the range of
angle close to $\pi$, and present the results in 
Fig. \ref{figBQ}
for $\epsn=0.01$. 
In case of volume $V \ll r^{3}_{0}$\cite{mig,kim}, 
$B_Q$ can be minimum at $\thH=\pi$ depending on the
ratio ${K}_2/{K}_1$ of the anisotropy constants.
$B_Q(\thH=\pi)$ is minimum for ${K}_2={K}_1$ while
it is maximum in the limit of ${K}_2 \ll {K}_1$
which is uniaxial symmetry\cite{mig}.
In the ferromagetic film 
which we have considered with small $\epsn$,
$B_Q$ has a minimum in the range of angle
$\pi/2 + O(\sqrt{\epsn}) < \thH < \pi - O(\sqrt{\epsn})$
for both ${K}_2={K}_1$ and ${K}_2 \ll {K}_1$,
as is shown in Table \ref{tab}.
Also, in the limit of ${K}_2 \gg {K}_1$ the
dependence of $f(\thH)$ on the ratio ${K}_2 /{K}_1$
is the same as that of $B_Q$ at $\thH=\pi$.
Thus, in the small $\epsn$ limit, 
the ratio does not make an effect on the shape
of $B_Q$, which is different from the situation for
$V \ll r^{3}_{0}$.\cite{kim}
In other words, since 
$B_Q$ has a shape similar to Fig. \ref{figBQ},
it always has a minimum in the
range of angle 
$\pi/2 + O(\sqrt{\epsn}) < \thH < \pi - O(\sqrt{\epsn})$.
The minimal value of $B_Q$ or
equivalently the maximal value
of $\Gamma_Q$ at $T=0$ can be obtained by the analysis
of $f(\thH)$ and its $\thH$ is given by
\ben
  \thH=\pi - \arctan [{2 \bar{K}_2 \over 
  \sqrt{(2 \bar{K}_2+{1 \over 2})^2+2} - {3 \over 2}} ]^{3/2},
\label{minth}
\een
where we notice that $\thH \approx 1.97(\simeq 112.9^{\circ})$
for, e.g.,  $K_1 = K_2$,
as is shown in Fig. \ref{figBQ}.
Simple analysis shows that the angle which gives the minimal
valus of $B_Q$, Eq. (\ref{minth}), decreases as 
$\bar{K}_2$ decreases, and converges to 
$\pi-\arctan (3\sqrt{3})( \approx 1.76 \simeq 
100.9^{\circ})$ in the 
limit of $\bar{K}_2=0$.\cite{mig}
Since the effective height of barrier $U^{eff}$ is
proportional to $\epsn$ for both
$\pi/2 + O(\sqrt{\epsn}) < \thH < \pi - O(\sqrt{\epsn})$
and $\thH=\pi$, we obtain $T_c \ \propto \epsn^{1/4}$ for
$\pi/2 + O(\sqrt{\epsn}) < \thH < \pi - O(\sqrt{\epsn})$
and $T_c \ \propto \epsn^{1/2}$ for
$\thH=\pi$. Since $T_c$ is inversely proportional
to $f(\thH)$ for a wide range of angles, 
$T_c$ becomes maximum at the angle (\ref{minth}).
For $K_1 \simeq K_2 \simeq 10^7 erg/cm^3$,
$M_0 \simeq 500 \ emu/cm^3$ and
$\epsn \sim 10^{-2}-10^{-3}$, the maximal value of $T_c$ 
is approximately $235-132 \ mK$ 
while $T_c \approx 167-53 \ mK$ for $\thH=\pi$.

In conclusion, we have studied the quantum nucleation
of a thin ferromagnetic film placed at some angle
with the magnetic field and presented the quantum
nucleation exponent and the
crossover temperature based on the spin
coherent path integral formalism.
If the experiment is to be performed, there are
three control parameters for comparison
with theory: the angle of the magnetic field $\thH$,
the magnitude of the field ${\bf H}$ in terms of
$\epsn$, and the temperature $T$. The dependence
of $B_Q$ and $T_c$ on the power of $\epsn$ 
seems to be observed by changing $\thH$. Also,
the angle (\ref{minth}) which
gives the maximal value of $T_c$ 
as well as $\Gamma_Q$
is expected to be observed in
experiment.

Discussions with D. S. Hwang are gratefully 
acknowledged.
This work was supported
in part by the Basic Science Research Institute Program,
Ministry of Education, Project No. BSRI-96-2414,
in part by the Ministry of Science and Technology of Korea
through the HTSRA, 
and in part by Non-Directed-Research-Fund,
Korea Research Foundation 1996.\\

\pagebreak
\begin{table}
\caption{ 
         Summary of the results for the quantum nucleation
         in the range of angle 
         $\pi/2 + O(\protect\sqrt{\epsn}) < \thH < 
         \pi - O(\protect\sqrt{\epsn})$
         and $\thH=\pi$. Here $B_Q$ is 
	 the WKB exponent at $T=0$ and $f(\thH)$  
	 given by Eq. (\protect\ref{fth}).}
 \label{tab}
\end{table}

\begin{center}
\begin{tabular}{|c|c|c|}  \hline
Field angle & 
$\pi/2 + O(\sqrt{\epsn}) < \thH < \pi - O(\sqrt{\epsn})$ & 
            $\thH=\pi$ \cite{chu}  \\ \hline
$\alpha=\beta$   & 1/2 & 1 \\
$M$   & $(2\bar{K}_2+\bar{H}_x / \sin \thc)^{-1}$
& $(4\bar{K}_2)^{-1}$ \\
$B_Q/J  r^{2}_{0}  z_t$    & $105.2 \ f(\thH)  \epsn^{3/4}$
& $37.9 \ \sqrt{K_1 / K_2} \ \epsn^{1/2}$ \\
$U^{eff}/K_1 r^{2}_{0}  z_t $ & 41.3 \epsn & 23.4 \epsn \\
$(k_B T_c) J/K_1$ & $0.22 \  \epsn^{1/4}/f(\thH)$ & 
$0.62 \ \sqrt{K_2/K_1} \epsn^{1/2}$ \\ 
\hline
\end{tabular}
\end{center}

\pagebreak

\begin{figure}
 \caption{Comparison of (a) $\sin (2 \theta_c)$ with
         (b) $\cos (2 \theta_c)$ in 
         Eqs. (\protect\ref{eq:etaeq})
         and (\protect\ref{eq:edelta}).
         Note that $\sin (2 \theta_c)=0$,
         $|\cos (2 \theta_c)|=1$ for both
         $\theta_H = \pi/2$ and $\pi$.}
 \label{figsth}
\end{figure}

\begin{figure}
 \caption{The shape of $\eta(=\thc-\thz)$ as a function
	  of $\thH$ for (a) $\epsn=0.01$ and
	  (b) $\epsn=0.001$. Note that 
	  $\eta \approx \protect\sqrt{2\epsn/3}$ is valid
	  in a wide range of angles, i.e., 
          $\pi/2 + O(\protect\sqrt{\epsn}) < \thH <
          \pi - O(\protect\sqrt{\epsn})$.}
 \label{etaval}
\end{figure}

\begin{figure}
 \caption{The $\thH$ dependence of the 
          quantum nucleation exponent $\bar{B}_Q$ for
          $K_1=K_2$ and $\epsn=0.01$,
	  where $\bar{B}_Q=B_Q/J  r^{2}_{0}  z_t$.
	  The minimal value of $\bar{B}_Q$ is approximately
	  2.65 at $\thH \approx 1.97$
          while $\bar{B}_Q \approx 3.79$ at $\thH=\pi$.}
 \label{figBQ}
\end{figure}

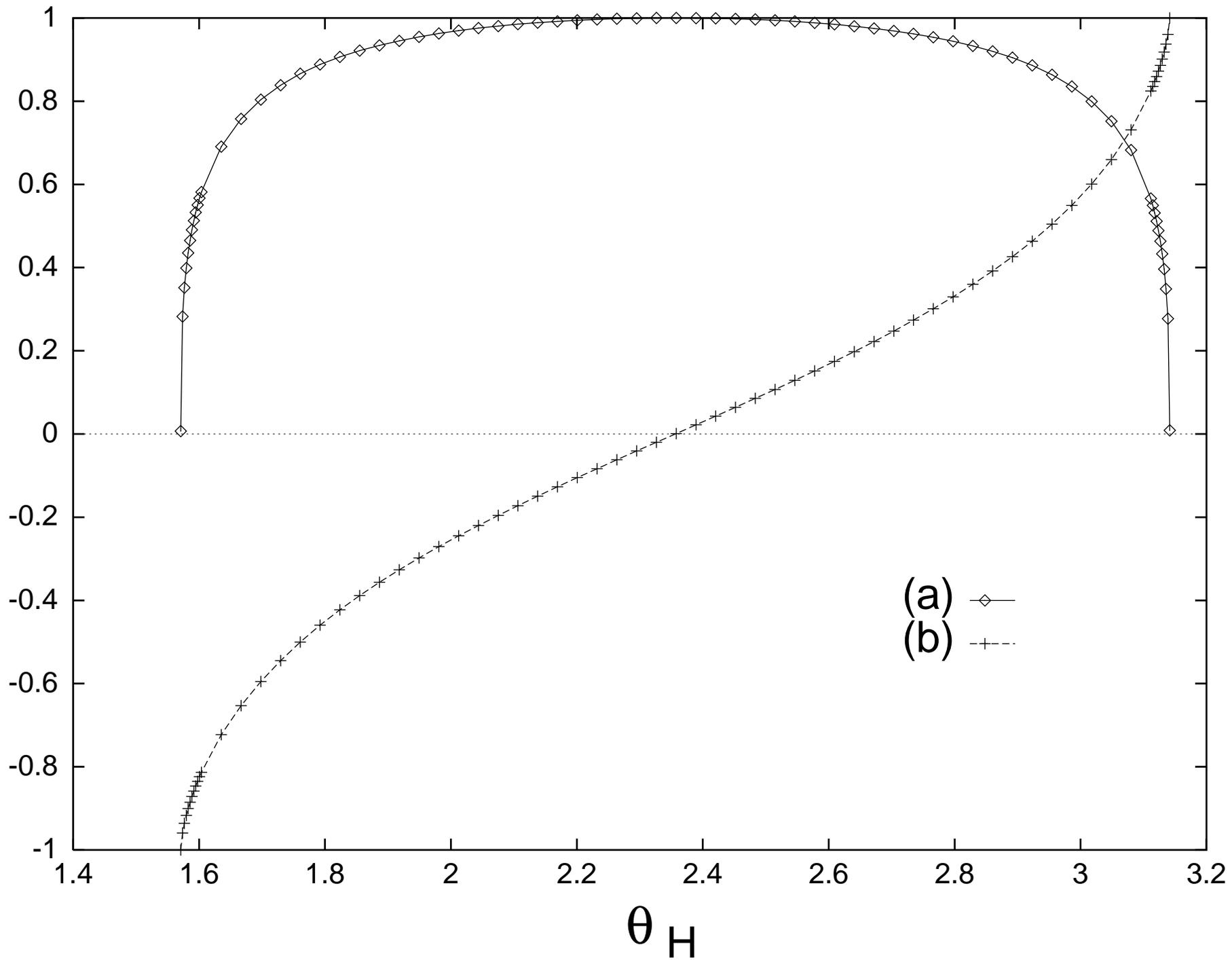
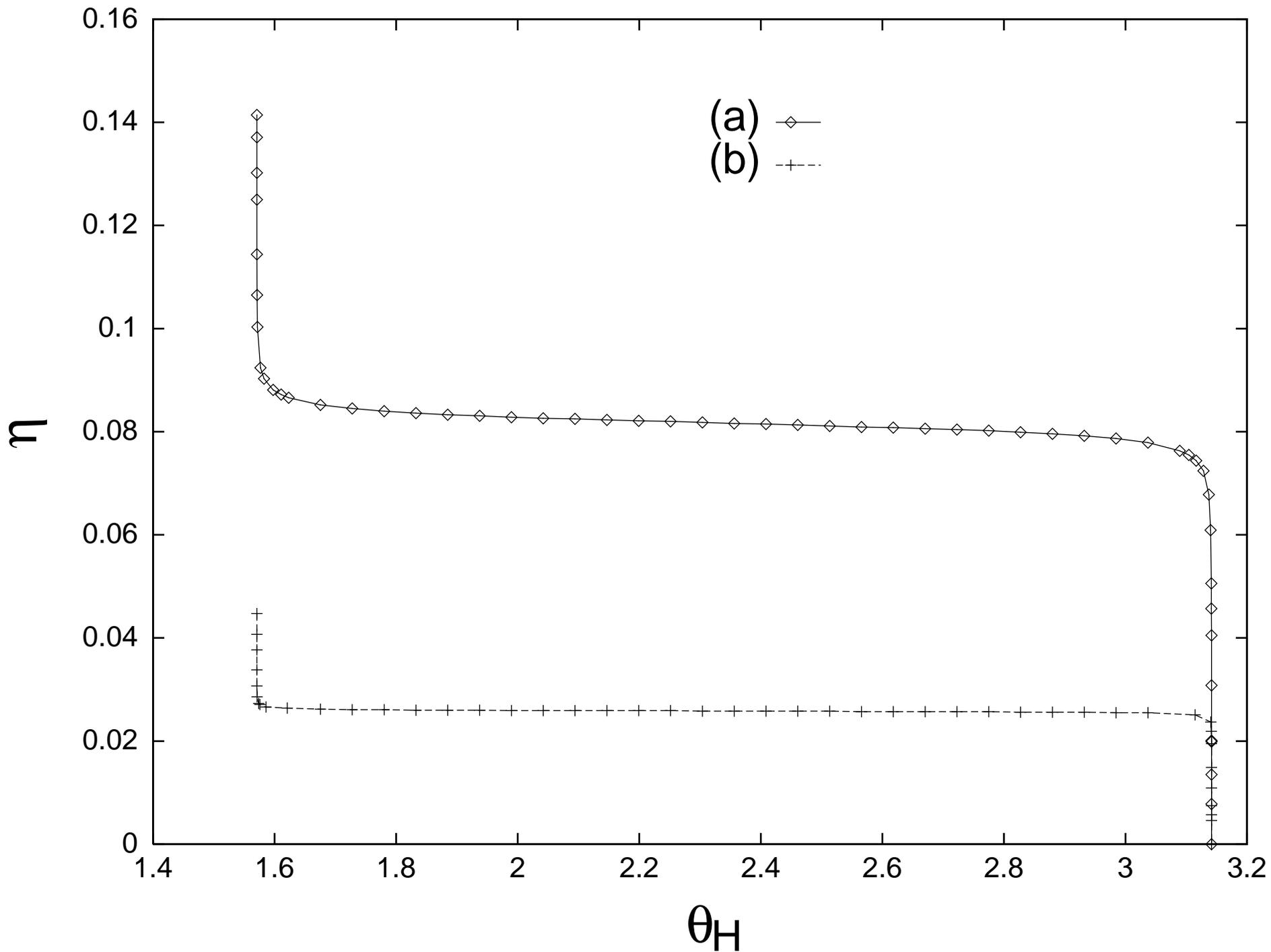
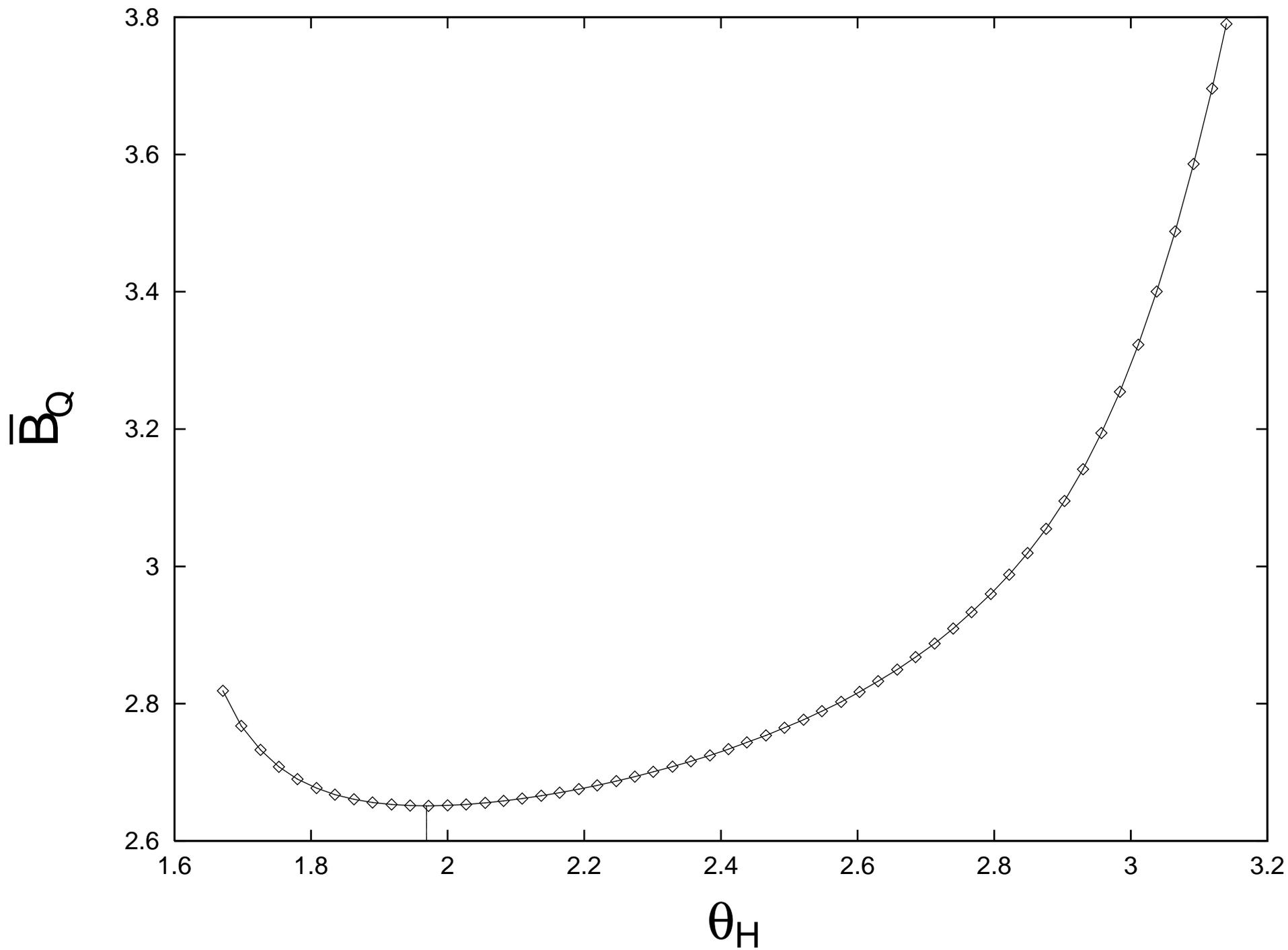
\end{document}